\documentclass[aps,prl,numerical,linenumbers,superscriptaddress,showpacs,floatfix,reprint]{revtex4-1}
\usepackage[dvipdfm]{graphicx}
\usepackage{amsmath,amssymb,amsfonts}
\usepackage{color}

\newcommand{\be}{\begin{eqnarray}}
\newcommand{\ee}{\end{eqnarray}}

\def\v2{\mbox{$v_2$}}

\begin{document}


%
\title{Is anisotropic flow really acoustic?}
%
%
%
\author{ Roy~A.~Lacey}
\email[E-mail: ]{Roy.Lacey@Stonybrook.edu}
\affiliation{Department of Chemistry, 
Stony Brook University, \\
Stony Brook, NY, 11794-3400, USA}
\author{Yi Gu} 
\affiliation{Department of Chemistry, 
Stony Brook University, \\
Stony Brook, NY, 11794-3400, USA}
\author{ X. Gong} 
\affiliation{Department of Chemistry, 
Stony Brook University, \\
Stony Brook, NY, 11794-3400, USA}
\author{D. Reynolds} 
\affiliation{Department of Chemistry, 
Stony Brook University, \\
Stony Brook, NY, 11794-3400, USA}
\author{ N.~N.~Ajitanand} 
\affiliation{Department of Chemistry, 
Stony Brook University, \\
Stony Brook, NY, 11794-3400, USA}
\author{ J.~M.~Alexander}
\affiliation{Department of Chemistry, 
Stony Brook University, \\
Stony Brook, NY, 11794-3400, USA}
\author{A. Mwai}
\affiliation{Department of Chemistry, 
Stony Brook University, \\
Stony Brook, NY, 11794-3400, USA}
\author{A.~Taranenko}
\affiliation{Department of Chemistry, 
Stony Brook University, \\
Stony Brook, NY, 11794-3400, USA} 
%



\date{\today}


\begin{abstract}

	The flow harmonics for charged hadrons ($v_{n}$) and their ratios $(v_n/v_2)_{n\geq 3}$, 
are studied for a broad range of transverse momenta ($p_T$) and centrality ($\text{cent}$) in Pb+Pb 
collisions at $\sqrt{s_{NN}}= 2.76$ TeV. They indicate characteristic scaling patterns for viscous damping 
consistent with the dispersion relation for sound propagation in the plasma produced in the collisions. 
These scaling properties are not only a unique signature for anisotropic expansion modulated by the 
specific shear viscosity ($\eta/s$), they provide essential constraints for the relaxation time, 
a distinction between two of the leading models for initial eccentricity, as well as an 
extracted $\left< \eta/s \right>$ value which is insensitive to the initial geometry model.
These constraints could be important for a more precise determination of $\eta/s$. 
	
\end{abstract}


\maketitle


Azimuthal anisotropy measurements are a key ingredient in ongoing efforts to 
pin down the precise value of the transport coefficients of the plasma 
produced in heavy ion collisions at both the Relativistic Heavy Ion Collider (RHIC) and the 
Large Hadron Collider (LHC). 
The Fourier coefficients $v_n$ are routinely used to quantify such measurements as a function 
of collision centrality (cent) and particle transverse momentum $p_T$;
\begin{equation}
\frac{dN}{d\phi} \propto \left( 1 + 2\sum_{n=1}v_n(p_T)\cos n(\phi - \psi_n) \right),
\label{eq:1}
\end{equation}  
where $\phi$ is the azimuthal angle of an emitted particle, 
and $\psi_{n}$ are the azimuths of the estimated participant event 
planes \cite{Ollitrault:1992bk,Adare:2010ux};
%
\begin{eqnarray}
v_n(p_T) = \left\langle \cos n(\phi - \psi_n) \right\rangle, \,\,\, \nonumber 
\label{eq:2}
\end{eqnarray}
where the brackets denote averaging over particles and events.
The distribution of the azimuthal angle difference 
($\Delta\phi =\phi_a - \phi_b$) between particle pairs with transverse 
momenta $p^a_{T}$ and $p^b_{T}$ (respectively) is also commonly used to 
quantify the anisotropy \cite{Lacey:2001va,ALICE:2011ab,Chatrchyan:2012wg,ATLAS:2012at};
%
%
\begin{equation}
\frac{dN^{\text{pairs}}}{d\Delta\phi} \propto \left( 1 + \sum_{n=1}2v_{n,n}(p^a_{T},p^b_{T})\cos(n\Delta\phi) \right),
\label{eq:3}
\end{equation}
\begin{eqnarray}
v_{n,n}(p^a_{T},p^b_{T}) = v_{n}(p^a_{T})v_{n}(p^b_{T}), \,\,\, \nonumber 
\label{eq:4}
\end{eqnarray}
where the latter factorization has been demonstrated to hold well for $p_T \alt 3$~GeV/c for 
particle pairs with a sizable pseudorapidity gap $\Delta\eta_p$ \cite{Chatrchyan:2012wg,ATLAS:2012at}.
 
The coefficients $v_{n}(p_T,\text{cent})$ (for $p_T \alt 3-4$~GeV/c) have been attributed to an eccentricity-driven 
hydrodynamic expansion of the plasma produced in the collision zone \cite{Heinz:2001xi,Teaney:2003kp,Huovinen:2001cy, 
Hirano:2002ds,Romatschke:2007mq,Song:2008hj,Schenke:2010rr}.
That is, a finite eccentricity moment $\varepsilon_n$ drives uneven pressure gradients in- and out 
of the event plane $\psi_n$, and the resulting expansion leads to the anisotropic flow of particles 
about this plane.  In this model framework, the values of $v_{n}(p_T,\text{cent})$ 
are sensitive to the magnitude of both $\varepsilon_n$ and the transport coefficient $\eta/s$
({\em i.e.}\ the specific shear viscosity or ratio of shear viscosity $\eta$ to 
entropy density $s$) of the expanding hot matter \cite{Heinz:2002rs,Teaney:2003kp,Lacey:2006pn,
Romatschke:2007mq,Drescher:2007cd,Xu:2007jv,Greco:2008fs}. Thus, $v_{n}(p_T,\text{cent})$ measurements 
provide a crucial bridge to the extraction of $\eta/s$ from data. 

Initial estimates of $\eta/s$ from $v_n$ measurements \cite{Lacey:2006bc,Adare:2006nq,Romatschke:2007mq,Luzum:2008cw,
Xu:2007jv,Drescher:2007cd,Song:2008hj,Lacey:2009xx,Dusling:2009df,Niemi:2012aj}
have all indicated a small value ($\eta/s \sim 1-4$ times the lower conjectured bound of ${1}/{4\pi}$ \cite{Kovtun:2004de}). 
Recent 3+1D hydrodynamic calculations, which have been quite successful 
at reproducing $v_{n}(p_T,\text{cent})$ measurements \cite{Schenke:2011bn,Gale:2012in,Gardim:2012yp}, have also 
indicated a similarly small value of $\eta/s \alt 2/4\pi$. However, the precision of all of these extractions has been 
hampered by significant theoretical uncertainty, especially those arising from poor constraints for the initial 
eccentricity and the relaxation time. One approach to the resolution of this issue is to target these uncertainties for 
systematic study, with the aim of establishing reliable upper and lower bounds for $\eta/s$ \cite{Song:2008hj,Luzum:2012wu}. 
An alternative approach, adopted in this work, is to ask whether better constraints for these theoretical bottlenecks 
can be developed to aid precision extractions of $\eta/s$?
%
\begin{figure*}[t]
\includegraphics[width=1.0\linewidth]{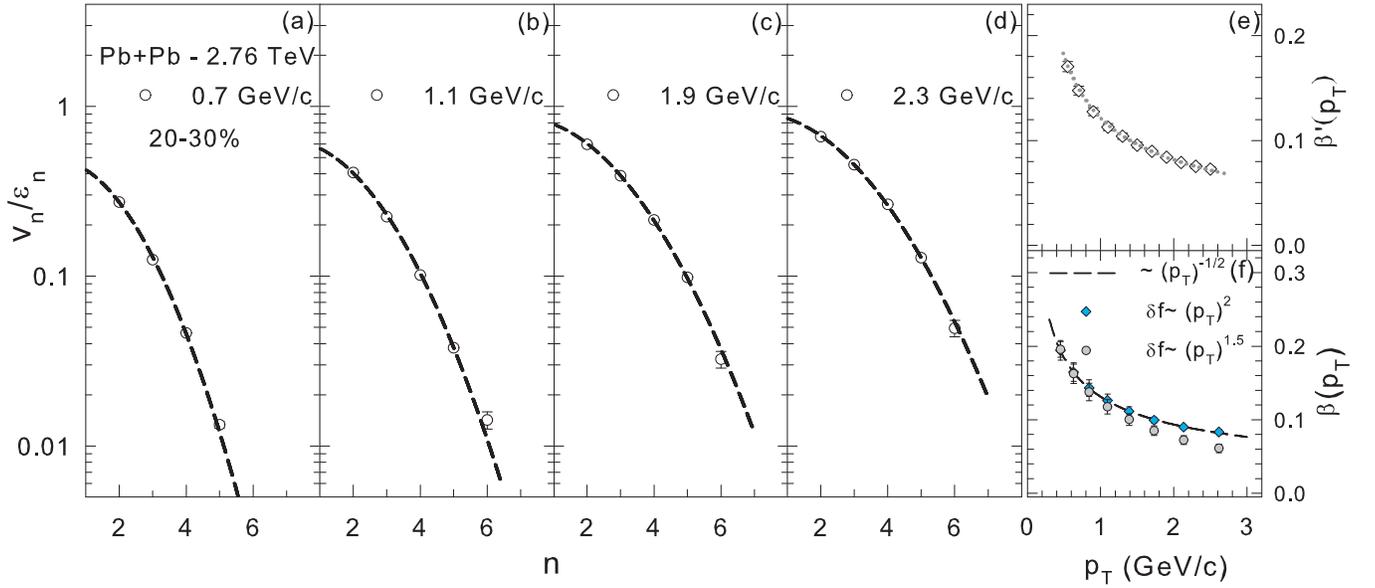}
\caption{(a)-(d) $v_{n}/\varepsilon_n$ vs. $n$ for charged hadrons for several $p_T$ selections in 20-30\% 
central Pb+Pb collisions at $\sqrt{s_{NN}}= 2.76$ TeV; (e) $\beta^{'}$ vs. $p_T$ for the same centrality selection;
(f) $\beta^{'}$ vs. $p_T$ from the analysis of the results from viscous hydrodynamical calculations \cite{Schenke:2011bn}
for $\delta f \propto p_T^{2}$ and $\delta f \propto p_T^{1.5}$.
The $v_n$ data are taken from Refs. \cite{ATLAS:2012at,JJia:2011hfa}; the dashed and dotted curves 
represent fits (see text).
}
\label{Fig1} 
\end{figure*}

	Given the acoustic nature of anisotropic flow ({\em i.e.} it is driven by pressure gradients),
a transparent way to evaluate the strength of the dissipative effects which reduce the magnitude 
of $v_{n}(p_T,\text{cent})$, is to consider the attenuation of sound waves in the plasma. 
In the presence of viscosity, sound intensity is exponentially damped $e^{(-r/\Gamma_s)}$ relative to the sound 
attenuation length $\Gamma_s$. This can be expressed as a perturbation to the energy-momentum 
tensor $T_{\mu\nu}$ \cite{Staig:2010pn};
\be
\delta T_{\mu\nu} (n,t) = \exp{\left(-\beta n^2 \right)}  \delta T_{\mu\nu} (0),
 \,\,\, \beta = \frac{2}{3} \frac{\eta}{s} \frac{1}{\bar{R}^2} \frac{t}{T}, 
\label{eq:5}
\ee 
which incorporates the dispersion relation for sound propagation, as well as the spectrum of 
initial ($t = 0$) perturbations associated with the eccentricity moments. The latter reflects 
the collision geometry and its associated density driven fluctuations. 
Here, the viscous coefficient $\beta \propto \eta/s$, $t \propto \bar{R}$ is the expansion time, 
$T$ is the temperature, $k = n/{\bar R}$ is the wave number ({\em i.e.} $2\pi {\bar R} = n\lambda$ 
for $n\ge 1$) and $\bar{R}$ is the transverse size of the collision zone.

The viscous corrections to $v_n$ implied in Eq. \ref{eq:5}, do not indicate an explicit $p_T$-dependence. 
However, a finite viscosity in the plasma results in an asymmetry in the energy-momentum 
tensor which manifests as a correction to the local particle 
distribution ($f$) at freeze-out \cite{Dusling:2009df};
\begin{equation}
    f = f_0 + \delta f(\tilde{p}_T), \,\,\, \tilde{p}_T = \frac{p_T}{T},
\label{eq:6}
\end{equation}
where $f_0$ is the equilibrium distribution and $\delta f(\tilde{p}_T)$ is its first order correction. 
The latter leads to the $p_T$-dependent viscous coefficient $\beta^{'}(\tilde{p}_T) \propto \beta/p_T^{\alpha}$, 
where the magnitude of $\alpha$ is related to the relaxation time $\tau_R(p_T)$.

Equations \ref{eq:5} and \ref{eq:6} suggest that for a given centrality, the viscous corrections to the  
flow harmonics $v_n(p_T)$, grow exponentially as $n^2$; 
\be
\frac{v_n(p_T)}{\varepsilon_n} \propto \exp{\left(-\beta^{'} \!n^2 \right)}, 
\label{eq:7}
\ee 
and the ratios $(v_n(p_T)/v_2(p_T))_{n\geq 3}$ can be expressed as;
\be
\frac{v_n(p_T)}{v_2(p_T)} = \frac{\varepsilon_n}{\varepsilon_2}\exp{\left(-\beta^{'}\!(n^2 - 4) \right)},  
\label{eq:8}
\ee 
indicating that they only depend on the eccentricity ratios and 
the relative viscous correction factors. Note as well that Eq.~\ref{eq:8} shows that the higher order 
harmonics $v_{n,n\geq 3}$, can all be expressed in terms of the lower order harmonic $v_2$, as has been 
observed recently \cite{ATLAS:2012at,Lacey:2011ug}. For a given harmonic, Eq.~\ref{eq:7} can be 
linearized to give 
\be
\ln\left(\frac{v_n(p_T)}{\varepsilon_n}\right) \propto \frac{-\beta^{''}}{\bar{R}}, 
\label{eq:9}
\ee 
which indicates a characteristic system size dependence ($1/\bar{R}$) of the viscous corrections.

If validated, the acoustic dissipative patterns summarized in Eqs. \ref{eq:7}, \ref{eq:8} and \ref{eq:9}, 
indicate that estimates for $\alpha$, $\beta$ and ${\varepsilon_n}/{\varepsilon_2}$ can be extracted directly 
from the data. Here, we perform validation tests for these dissipative patterns with an eye toward 
more stringent constraints for $\tau_R$, $\eta/s$ and the distinction between different 
eccentricity models. 

	The data employed in our analysis are taken from measurements by the ATLAS 
collaboration for Pb+Pb collisions at $\sqrt{s_{NN}}$ = 2.76 TeV \cite{ATLAS:2012at,JJia:2011hfa}. 
These measurements exploit the event plane analysis method (c.f. Eq. \ref{eq:1}), as well as the  
two-particle $\Delta\phi$ correlation technique (c.f. Eq. \ref{eq:3}) to obtain 
robust values of $v_{n}(p_T,\text{cent})$ for a sizable  $\Delta\eta_p $ gap between particles 
and the event plane, or particle pairs. 
%
\begin{figure*}[t]
\includegraphics[width=1.0\linewidth]{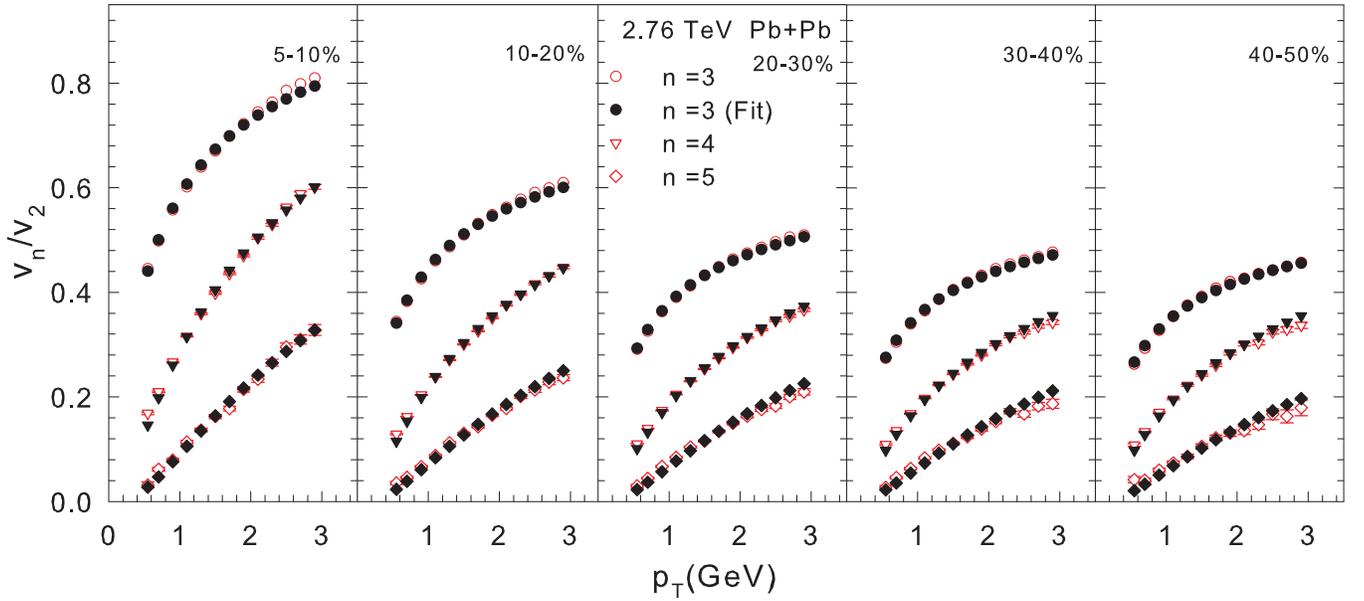}
%
\caption{$v_n/v_2$ vs. $p_T$ for several centrality selections for Pb+Pb collisions at $\sqrt{s_{NN}}= 2.76$ TeV. 
The open symbols show the values obtained from data; the filled symbols show the results of fits to these ratios 
with Eq. \ref{eq:8} (see text).
}
\label{Fig2}
\end{figure*}
We divide these values by $\varepsilon_n({\text{cent}})$ and plot them as a function of $n$,
to make an initial test for viscous damping compatible with sound propagation in the plasma produced 
in these collisions. Monte Carlo Glauber (MC-Glauber) simulations were used to compute 
the number of participants N$_\text{part}(\text{cent})$ and $\varepsilon_n({\text{cent}})$ from 
the two-dimensional profile of the density of sources in the transverse  plane $\rho_s(\mathbf{r_{\perp}})$.   
The weight $\omega(\mathbf{r_{\perp}}) = \mathbf{r_{\perp}\!}^n$ \cite{Lacey:2010hw} was used to 
compute $\varepsilon_n({\text{cent}})$. 

The open circles in Figs.~\ref{Fig1} (a)-(d) show representative examples of $v_n/\varepsilon_n$ vs. $n$ for 
several $p_T$ cuts, for the 20-30\% centrality selection. The dashed curves which indicate 
fits to the data with Eq. \ref{eq:7}, confirm the expected exponential growth of the viscous 
corrections to $v_n$, as $n^2$. The $p_T$-dependent viscous coefficients $\beta^{'}(\tilde{p}_T)$ obtained from 
these fits, are summarized in Fig.~\ref{Fig1}~(e); they show the expected $1/p_T^{\alpha}$ dependence 
attributable to $\delta f(p_T)$. Note that a similar dependence is obtained for fits to the results of 
viscous hydrodynamical calculations, as illustrated in panel (f). The latter indicates that the 
$p_T$ dependence of $\beta$ allows a distinction between the two sets of calculations which use 
different input assumptions for $\delta f(p_T)$. The dotted curve in panel (e) is a fit which gives 
the values $\alpha \sim 0.58$ and $\beta \sim 0.12$.  Similar results were obtained for a broad 
range of centrality selections.

Additional constraints can be obtained from the ratios of the flow harmonics $(v_n(p_T)/v_2(p_T))_{n\ge 3}$ (cf. Eq.~\ref{eq:8}), 
as well as the dependence of $v_n(p_T)/\varepsilon_n$ on the transverse size of the collision zone (cf. Eq.~\ref{eq:9}). 
The open symbols in Fig. \ref{Fig2} show the values of $(v_n(p_T)/v_2(p_T))$ for $n=3,4$ and $5$, for each of 
the centrality selections indicated. A simultaneous fit to these ratios 
was performed with Eq.~\ref{eq:8} to extract $\beta$ and ${\varepsilon_n}/{\varepsilon_2}$ at 
each centrality. 
Small variations about the previously extracted value of $\alpha \sim 0.58$ were used to 
aid the convergence of these fits.
The filled symbols in Fig. \ref{Fig2} show the excellent fits achieved; they confirm 
the characteristic dependence of the relative viscous correction factors expressed in Eq.~\ref{eq:8}. 
They also confirm that the relationship between $v_2$ and the higher order 
harmonics stems solely from ``acoustic scaling'' of the viscous corrections to anisotropic flow. 
The extracted values for ${\varepsilon_n}/{\varepsilon_2}$, $\alpha$ and $\beta$ are summarized 
and discussed below.

\begin{figure*}[t]
\begin{tabular}{cc}
\includegraphics[width=0.54\linewidth]{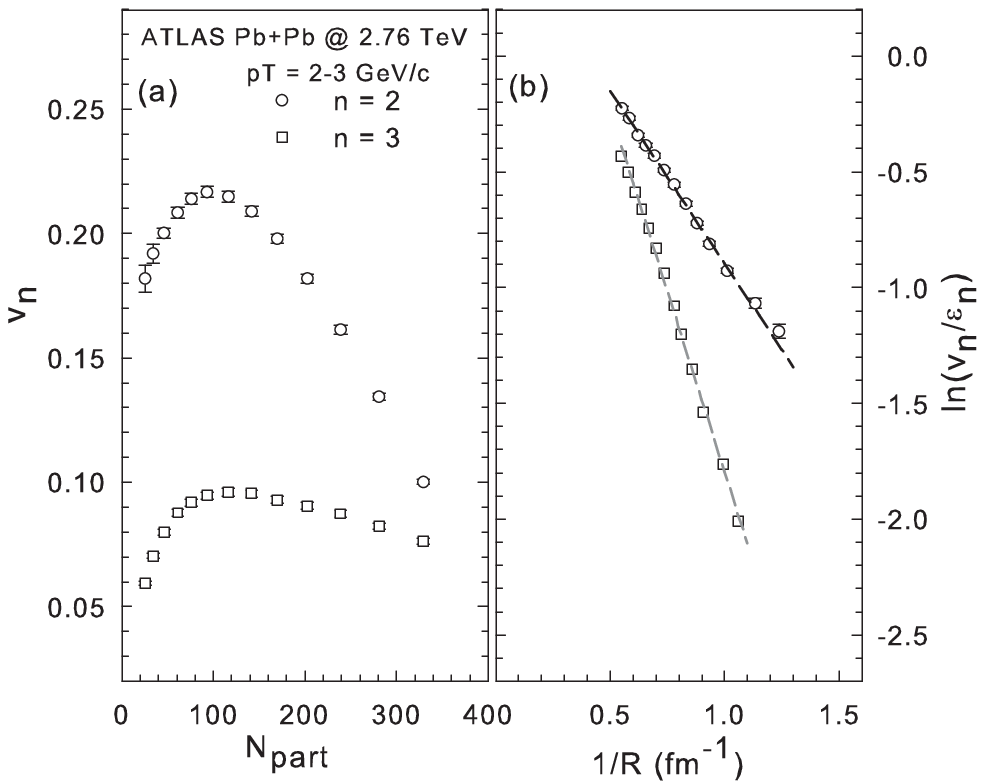} 
\includegraphics[width=0.46\linewidth]{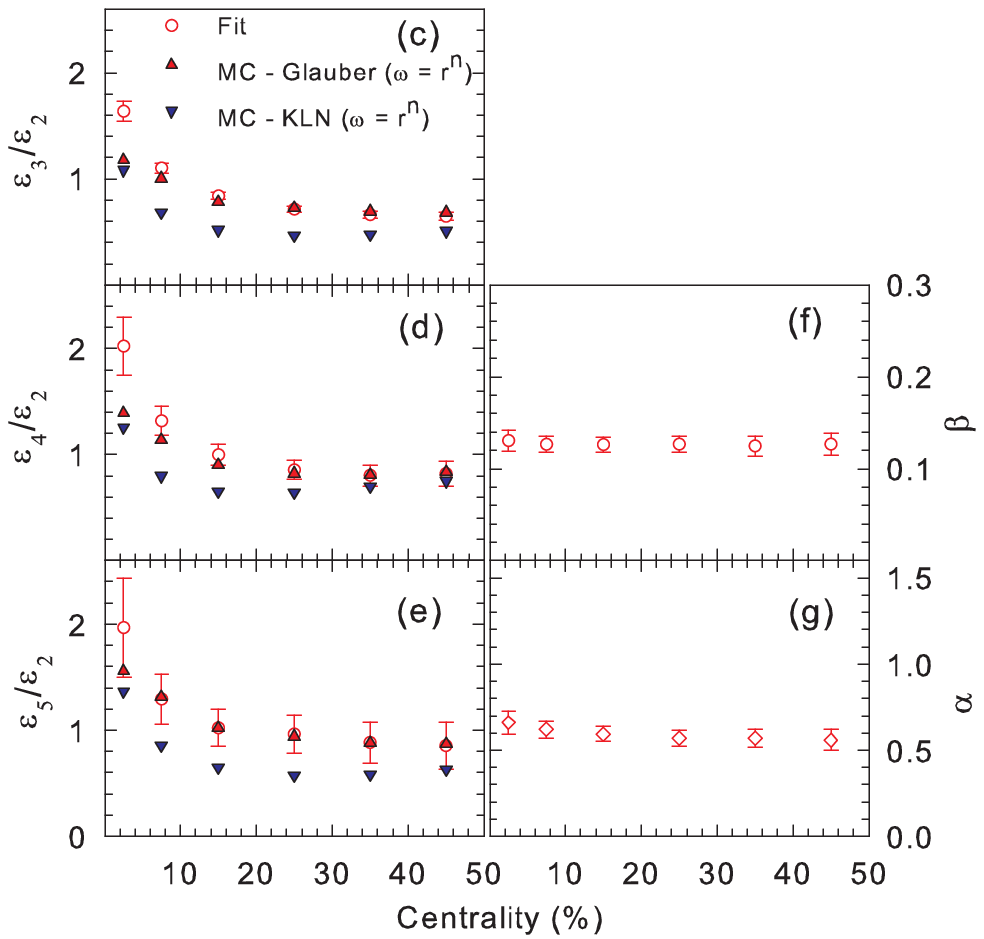}
\end{tabular}
\caption{(a) $v_{2,3}$ vs. N$_\text{part}$ for $p_T = 1-2$ GeV/c: 
(b) $\ln(v_n/\varepsilon_n)$ vs. $1/\bar{R}$ for the 
data shown in (a): (c - e) centrality dependence of the $\varepsilon_n/\varepsilon_2$ ratios 
extracted from fits to $(v_n(p_T)/v_2(p_T))_{n\ge 3}$ with Eq.~\ref{eq:8}; 
$\varepsilon_n/\varepsilon_2$ ratios for the MC-Glauber \cite{Miller:2007ri,Lacey:2010hw} 
and MC-KLN \cite{Drescher:2007ax} models are also shown: (f) extracted values of $\beta$ vs. centrality: 
(g) extracted values of $\alpha$ vs. centrality (see text).
}
\label{Fig3}
\end{figure*}

Figures~\ref{Fig3}(a) and (b) gives a more transparent view of the influence of system 
size on the viscous corrections. Fig.~\ref{Fig3}(a) shows that $v_{2,3}$ increases 
for $140 \alt \text{N}_{\text{part}} \alt 340$ as would be expected from an increase in 
$\varepsilon_{2,3}$ over the same $\text{N}_{\text{part}}$ range. For $\text{N}_{\text{part}} \alt 140$ however,
the decreasing trend of $v_{2,3}$ contrasts with the increasing trends for $\varepsilon_{2,3}$, suggesting 
that the viscous effects due to much smaller system sizes, serve to suppress $v_{2,3}$. This is confirmed by
the dashed curves in Fig.~\ref{Fig3}(b) which validate the expected linear dependence of 
$\ln(v_n/\varepsilon_n)$ on $1/\bar{R}$  (cf. Eq.~\ref{eq:9}) for the data shown in Fig.~\ref{Fig3}(a).  
A similar dependence was observed for other $p_T$ selections.
The slopes of these curves serve as an important additional constraint for $\beta$.

 Figures \ref{Fig3}(c) - (e) show a comparison between the  
${\varepsilon_n}/{\varepsilon_2}$ ratios extracted from the fits shown in 
Fig.~\ref{Fig2} (open symbols), and those obtained from 
model calculations (filled symbols). For the 5-50\% centrality range, the comparison shows 
good agreement between the extracted ratios and those obtained from MC-Glauber calculations with 
weight $\omega(\mathbf{r_{\perp}}) = \mathbf{r_{\perp}\!}^n$ \cite{Lacey:2010hw}.
A similarly good agreement with the ratios obtained from a Monte Carlo 
implementation \cite{Drescher:2007ax} of the factorized 
Kharzeev-Levin-Nardi (KLN) model \cite{Kharzeev:2000ph,Lappi:2006xc} 
is not observed. For the 0-5\% most central collisions, the extracted values of 
${\varepsilon_n}/{\varepsilon_2}$ are larger than the values obtained from either 
eccentricity model. This difference could result from an overestimate of 
$\varepsilon_2$ in the 0-5\% centrality selection, for the initial eccentricity 
models considered. 
%
%
%
%
\begin{figure*}[t]
\vskip 0.3cm
\includegraphics[width=0.75\linewidth]{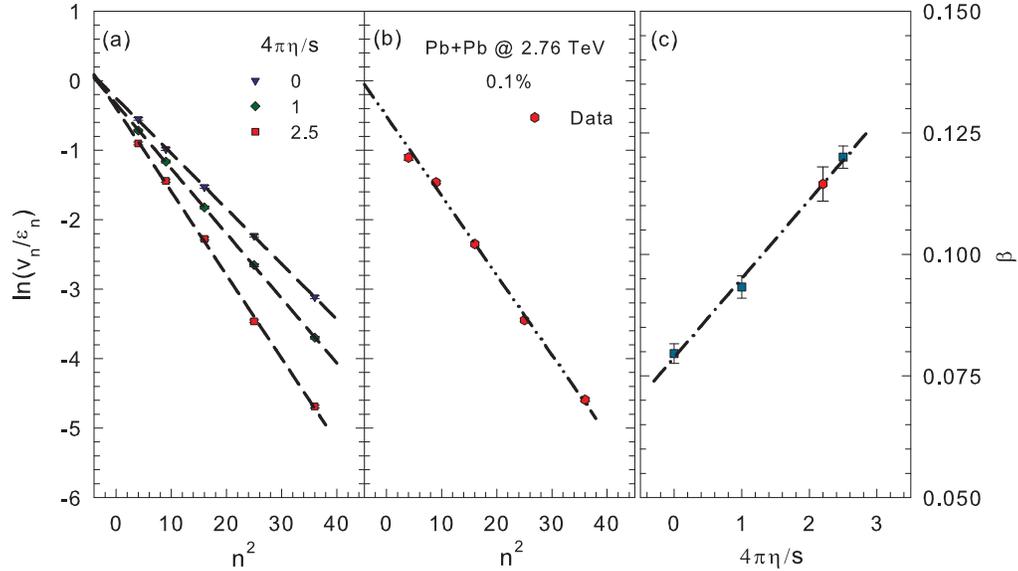}
\caption{(a) $\ln(v_n/\varepsilon_n)$ vs. $n^2$ from viscous hydrodynamical calculations for 
three values of specific shear viscosity as indicated. (b) $\ln(v_n/\varepsilon_n)$ vs. $n^2$ for Pb+Pb data.
The $p_T$-integrated $v_n$ results in (a) and (b) are for 0.1\% central Pb+Pb collisions
at $\sqrt{s_{NN}}= 2.76$ TeV \cite{cms_ulc_note}; the curves are linear fits. 
(c) $\beta$ vs. $4\pi\eta/s$ extracted from the curves shown in (a) and (b).
}
\label{Fig4}
\end{figure*}

The fits shown in Fig.~\ref{Fig2} also give values for $\alpha$ and $\beta$, which are 
summarized in Figs.~\ref{Fig3}(f) and (g); they are essentially independent of centrality. 
This suggests that, within errors, the full data set 
for $v_{n}(p_T,\text{cent})$ can be understood in terms of the eccentricity moments 
coupled to a single (average) value for $\alpha$ and $\beta$ (respectively). This observation is 
compatible with recent viscous hydrodynamical calculations which have been successful in
reproducing $v_{n}(p_T,\text{cent})$ measurements with a single $\delta f(\tilde{p}_T)$ 
ansatz and an average value of $\eta/s$ \cite{Schenke:2011bn,Gale:2012in}. Therefore, these values 
of $\alpha$ and $\beta$ should provide an important set of constraints for detailed 
model calculations. 
	
	To demonstrate their utility, we have used the results from recent viscous 
hydrodynamical calculations \cite{cms_ulc_note} to calibrate $\beta$ and make 
an estimate of $\eta/s$. This is illustrated in Fig.~\ref{Fig4}. The $p_T$-integrated 
$v_n$ results from viscous hydrodynamical calculations for three separate $\eta/s$ values, 
for 0.1\% central Pb+Pb collisions are shown in Fig.~\ref{Fig4}(a). They indicate the expected 
linear dependence of $\ln(v_n/\epsilon_n)$ on $n^2$, as well as the required sensitivity of 
the slopes of these curves to the magnitude of $\eta/s$. The calibration curve or $\beta$ vs. $4\pi\eta/s$, obtained 
from linear fits to the curves in Fig.~\ref{Fig4}(a), is shown in Fig.~\ref{Fig4}(c). 
The $p_T$-integrated $v_n$ data \cite{cms_ulc_note} shown in Fig.~\ref{Fig4}(b), also validates  
the expected linear dependence of $\ln(v_n/\epsilon_n)$ on $n^2$ for the same $\varepsilon_n$ 
values employed in Fig.~\ref{Fig4}(a). We use the slope of this curve in concert with the 
calibration in Fig.~\ref{Fig4}(c) to obtain the estimate $\left< 4\pi\eta/s \right> \sim 2.2 \pm 0.2$,  
which is in reasonable agreement with recent $\left< \eta/s \right>$
estimates \cite{Lacey:2011ug,Schenke:2011tv,Qiu:2011hf,Schenke:2011bn,Gale:2012in}.
Here, it is noteworthy that our calibration procedure leads to a $\left< \eta/s \right>$ value  
which is insensitive to the initial geometry model employed.   
Further calculations are undoubtedly required to reduce model driven calibration 
uncertainties. However, our analysis clearly demonstrates the value of the relative 
magnitudes of $v_n$ as an important constraint.


In summary, we have presented a detailed phenomenological study of viscous damping of the flow 
harmonics $v_n$ and their ratios $(v_n/(v_2))_{n\geq 3}$, for 
Pb+Pb collisions at $\sqrt{s_{NN}}= 2.76$ TeV. 
Within a parametrized viscous hydrodynamical framework,
this damping can be understood to be a consequence of the acoustic nature of 
anisotropic flow. That is, the observed viscous damping reflects the detailed scaling 
properties inferred from the dispersion relation for sound propagation in the plasma 
produced in these collisions. These patterns give a unique signature for anisotropic 
expansion modulated by viscosity, and provide straightforward constraints for 
the relaxation time, a distinction between two of the leading models for 
initial eccentricity, as well as an extracted $\left< \eta/s \right>$ value which is 
essentially independent of the initial eccentricity. Such constraints could be crucial 
for a more precise determination of the specific shear viscosity $\eta/s$.

{\bf Acknowledgments}
This research is supported by the US DOE under contract DE-FG02-87ER40331.A008. 
 


%
\bibliography{vn_scaling_LHC2} 

\begin{thebibliography}{40}%
\makeatletter
\providecommand \@ifxundefined [1]{%
 \@ifx{#1\undefined}
}%
\providecommand \@ifnum [1]{%
 \ifnum #1\expandafter \@firstoftwo
 \else \expandafter \@secondoftwo
 \fi
}%
\providecommand \@ifx [1]{%
 \ifx #1\expandafter \@firstoftwo
 \else \expandafter \@secondoftwo
 \fi
}%
\providecommand \natexlab [1]{#1}%
\providecommand \enquote  [1]{``#1''}%
\providecommand \bibnamefont  [1]{#1}%
\providecommand \bibfnamefont [1]{#1}%
\providecommand \citenamefont [1]{#1}%
\providecommand \href@noop [0]{\@secondoftwo}%
\providecommand \href [0]{\begingroup \@sanitize@url \@href}%
\providecommand \@href[1]{\@@startlink{#1}\@@href}%
\providecommand \@@href[1]{\endgroup#1\@@endlink}%
\providecommand \@sanitize@url [0]{\catcode `\\12\catcode `\$12\catcode
  `\&12\catcode `\#12\catcode `\^12\catcode `\_12\catcode `\%12\relax}%
\providecommand \@@startlink[1]{}%
\providecommand \@@endlink[0]{}%
\providecommand \url  [0]{\begingroup\@sanitize@url \@url }%
\providecommand \@url [1]{\endgroup\@href {#1}{\urlprefix }}%
\providecommand \urlprefix  [0]{URL }%
\providecommand \Eprint [0]{\href }%
\providecommand \doibase [0]{http://dx.doi.org/}%
\providecommand \selectlanguage [0]{\@gobble}%
\providecommand \bibinfo  [0]{\@secondoftwo}%
\providecommand \bibfield  [0]{\@secondoftwo}%
\providecommand \translation [1]{[#1]}%
\providecommand \BibitemOpen [0]{}%
\providecommand \bibitemStop [0]{}%
\providecommand \bibitemNoStop [0]{.\EOS\space}%
\providecommand \EOS [0]{\spacefactor3000\relax}%
\providecommand \BibitemShut  [1]{\csname bibitem#1\endcsname}%
\let\auto@bib@innerbib\@empty
\bibitem [{\citenamefont {Ollitrault}(1992)}]{Ollitrault:1992bk}%
  \BibitemOpen
  \bibfield  {author} {\bibinfo {author} {\bibfnamefont {J.-Y.}\ \bibnamefont
  {Ollitrault}},\ }\href@noop {} {\bibfield  {journal} {\bibinfo  {journal}
  {Phys. Rev.}\ }\textbf {\bibinfo {volume} {D46}},\ \bibinfo {pages} {229}
  (\bibinfo {year} {1992})}\BibitemShut {NoStop}%
\bibitem [{\citenamefont {Adare}\ \emph {et~al.}(2010)\citenamefont {Adare}
  \emph {et~al.}}]{Adare:2010ux}%
  \BibitemOpen
  \bibfield  {author} {\bibinfo {author} {\bibfnamefont {A.}~\bibnamefont
  {Adare}} \emph {et~al.} (\bibinfo {collaboration} {PHENIX}),\ }\href
  {\doibase 10.1103/PhysRevLett.105.062301} {\bibfield  {journal} {\bibinfo
  {journal} {Phys. Rev. Lett.}\ }\textbf {\bibinfo {volume} {105}},\ \bibinfo
  {pages} {062301} (\bibinfo {year} {2010})},\ \Eprint
  {http://arxiv.org/abs/1003.5586} {arXiv:1003.5586 [nucl-ex]} \BibitemShut
  {NoStop}%
\bibitem [{\citenamefont {Lacey}(2002)}]{Lacey:2001va}%
  \BibitemOpen
  \bibfield  {author} {\bibinfo {author} {\bibfnamefont {R.~A.}\ \bibnamefont
  {Lacey}},\ }\href@noop {} {\bibfield  {journal} {\bibinfo  {journal} {Nucl.
  Phys.}\ }\textbf {\bibinfo {volume} {A698}},\ \bibinfo {pages} {559}
  (\bibinfo {year} {2002})}\BibitemShut {NoStop}%
\bibitem [{\citenamefont {Aamodt}\ \emph {et~al.}(2011)\citenamefont {Aamodt}
  \emph {et~al.}}]{ALICE:2011ab}%
  \BibitemOpen
  \bibfield  {author} {\bibinfo {author} {\bibfnamefont {K.}~\bibnamefont
  {Aamodt}} \emph {et~al.} (\bibinfo {collaboration} {ALICE Collaboration}),\
  }\href {\doibase 10.1103/PhysRevLett.107.032301} {\bibfield  {journal}
  {\bibinfo  {journal} {Phys.Rev.Lett.}\ }\textbf {\bibinfo {volume} {107}},\
  \bibinfo {pages} {032301} (\bibinfo {year} {2011})},\ \Eprint
  {http://arxiv.org/abs/1105.3865} {arXiv:1105.3865 [nucl-ex]} \BibitemShut
  {NoStop}%
\bibitem [{\citenamefont {Chatrchyan}\ \emph {et~al.}(2012)\citenamefont
  {Chatrchyan} \emph {et~al.}}]{Chatrchyan:2012wg}%
  \BibitemOpen
  \bibfield  {author} {\bibinfo {author} {\bibfnamefont {S.}~\bibnamefont
  {Chatrchyan}} \emph {et~al.} (\bibinfo {collaboration} {CMS Collaboration}),\
  }\href {\doibase 10.1140/epjc/s10052-012-2012-3} {\bibfield  {journal}
  {\bibinfo  {journal} {Eur.Phys.J.}\ }\textbf {\bibinfo {volume} {C72}},\
  \bibinfo {pages} {2012} (\bibinfo {year} {2012})},\ \Eprint
  {http://arxiv.org/abs/1201.3158} {arXiv:1201.3158 [nucl-ex]} \BibitemShut
  {NoStop}%
\bibitem [{\citenamefont {Aad}\ \emph {et~al.}(2012)\citenamefont {Aad} \emph
  {et~al.}}]{ATLAS:2012at}%
  \BibitemOpen
  \bibfield  {author} {\bibinfo {author} {\bibfnamefont {G.}~\bibnamefont
  {Aad}} \emph {et~al.} (\bibinfo {collaboration} {ATLAS Collaboration}),\
  }\href {\doibase 10.1103/PhysRevC.86.014907} {\bibfield  {journal} {\bibinfo
  {journal} {Phys.Rev.}\ }\textbf {\bibinfo {volume} {C86}},\ \bibinfo {pages}
  {014907} (\bibinfo {year} {2012})},\ \Eprint {http://arxiv.org/abs/1203.3087}
  {arXiv:1203.3087 [hep-ex]} \BibitemShut {NoStop}%
\bibitem [{\citenamefont {Heinz}\ and\ \citenamefont
  {Kolb}(2002)}]{Heinz:2001xi}%
  \BibitemOpen
  \bibfield  {author} {\bibinfo {author} {\bibfnamefont {U.}~\bibnamefont
  {Heinz}}\ and\ \bibinfo {author} {\bibfnamefont {P.}~\bibnamefont {Kolb}},\
  }\href@noop {} {\bibfield  {journal} {\bibinfo  {journal} {Nucl. Phys.}\
  }\textbf {\bibinfo {volume} {A702}},\ \bibinfo {pages} {269} (\bibinfo {year}
  {2002})}\BibitemShut {NoStop}%
\bibitem [{\citenamefont {Teaney}(2003)}]{Teaney:2003kp}%
  \BibitemOpen
  \bibfield  {author} {\bibinfo {author} {\bibfnamefont {D.}~\bibnamefont
  {Teaney}},\ }\href {\doibase 10.1103/PhysRevC.68.034913} {\bibfield
  {journal} {\bibinfo  {journal} {Phys. Rev.}\ }\textbf {\bibinfo {volume}
  {C68}},\ \bibinfo {pages} {034913} (\bibinfo {year} {2003})}\BibitemShut
  {NoStop}%
\bibitem [{\citenamefont {Huovinen}\ \emph {et~al.}(2001)\citenamefont
  {Huovinen}, \citenamefont {Kolb}, \citenamefont {Heinz}, \citenamefont
  {Ruuskanen},\ and\ \citenamefont {Voloshin}}]{Huovinen:2001cy}%
  \BibitemOpen
  \bibfield  {author} {\bibinfo {author} {\bibfnamefont {P.}~\bibnamefont
  {Huovinen}}, \bibinfo {author} {\bibfnamefont {P.~F.}\ \bibnamefont {Kolb}},
  \bibinfo {author} {\bibfnamefont {U.~W.}\ \bibnamefont {Heinz}}, \bibinfo
  {author} {\bibfnamefont {P.~V.}\ \bibnamefont {Ruuskanen}}, \ and\ \bibinfo
  {author} {\bibfnamefont {S.~A.}\ \bibnamefont {Voloshin}},\ }\href@noop {}
  {\bibfield  {journal} {\bibinfo  {journal} {Phys. Lett.}\ }\textbf {\bibinfo
  {volume} {B503}},\ \bibinfo {pages} {58} (\bibinfo {year}
  {2001})}\BibitemShut {NoStop}%
\bibitem [{\citenamefont {Hirano}\ and\ \citenamefont
  {Tsuda}(2002)}]{Hirano:2002ds}%
  \BibitemOpen
  \bibfield  {author} {\bibinfo {author} {\bibfnamefont {T.}~\bibnamefont
  {Hirano}}\ and\ \bibinfo {author} {\bibfnamefont {K.}~\bibnamefont {Tsuda}},\
  }\href {\doibase 10.1103/PhysRevC.66.054905} {\bibfield  {journal} {\bibinfo
  {journal} {Phys. Rev.}\ }\textbf {\bibinfo {volume} {C66}},\ \bibinfo {pages}
  {054905} (\bibinfo {year} {2002})},\ \Eprint
  {http://arxiv.org/abs/nucl-th/0205043} {arXiv:nucl-th/0205043} \BibitemShut
  {NoStop}%
\bibitem [{\citenamefont {Romatschke}\ and\ \citenamefont
  {Romatschke}(2007)}]{Romatschke:2007mq}%
  \BibitemOpen
  \bibfield  {author} {\bibinfo {author} {\bibfnamefont {P.}~\bibnamefont
  {Romatschke}}\ and\ \bibinfo {author} {\bibfnamefont {U.}~\bibnamefont
  {Romatschke}},\ }\href {\doibase 10.1103/PhysRevLett.99.172301} {\bibfield
  {journal} {\bibinfo  {journal} {Phys. Rev. Lett.}\ }\textbf {\bibinfo
  {volume} {99}},\ \bibinfo {pages} {172301} (\bibinfo {year}
  {2007})}\BibitemShut {NoStop}%
\bibitem [{\citenamefont {Song}\ and\ \citenamefont
  {Heinz}(2009)}]{Song:2008hj}%
  \BibitemOpen
  \bibfield  {author} {\bibinfo {author} {\bibfnamefont {H.}~\bibnamefont
  {Song}}\ and\ \bibinfo {author} {\bibfnamefont {U.~W.}\ \bibnamefont
  {Heinz}},\ }\href {\doibase 10.1088/0954-3899/36/6/064033} {\bibfield
  {journal} {\bibinfo  {journal} {J. Phys.}\ }\textbf {\bibinfo {volume}
  {G36}},\ \bibinfo {pages} {064033} (\bibinfo {year} {2009})}\BibitemShut
  {NoStop}%
\bibitem [{\citenamefont {Schenke}\ \emph {et~al.}(2010)\citenamefont
  {Schenke}, \citenamefont {Jeon},\ and\ \citenamefont
  {Gale}}]{Schenke:2010rr}%
  \BibitemOpen
  \bibfield  {author} {\bibinfo {author} {\bibfnamefont {B.}~\bibnamefont
  {Schenke}}, \bibinfo {author} {\bibfnamefont {S.}~\bibnamefont {Jeon}}, \
  and\ \bibinfo {author} {\bibfnamefont {C.}~\bibnamefont {Gale}},\ }\href@noop
  {} {\  (\bibinfo {year} {2010})},\ \Eprint {http://arxiv.org/abs/1009.3244}
  {arXiv:1009.3244 [hep-ph]} \BibitemShut {NoStop}%
\bibitem [{\citenamefont {Heinz}\ and\ \citenamefont
  {Wong}(2002)}]{Heinz:2002rs}%
  \BibitemOpen
  \bibfield  {author} {\bibinfo {author} {\bibfnamefont {U.~W.}\ \bibnamefont
  {Heinz}}\ and\ \bibinfo {author} {\bibfnamefont {S.~M.~H.}\ \bibnamefont
  {Wong}},\ }\href {\doibase 10.1103/PhysRevC.66.014907} {\bibfield  {journal}
  {\bibinfo  {journal} {Phys. Rev.}\ }\textbf {\bibinfo {volume} {C66}},\
  \bibinfo {pages} {014907} (\bibinfo {year} {2002})}\BibitemShut {NoStop}%
\bibitem [{\citenamefont {Lacey}\ and\ \citenamefont
  {Taranenko}(2006)}]{Lacey:2006pn}%
  \BibitemOpen
  \bibfield  {author} {\bibinfo {author} {\bibfnamefont {R.~A.}\ \bibnamefont
  {Lacey}}\ and\ \bibinfo {author} {\bibfnamefont {A.}~\bibnamefont
  {Taranenko}},\ }\href@noop {} {\bibfield  {journal} {\bibinfo  {journal}
  {PoS}\ }\textbf {\bibinfo {volume} {CFRNC2006}},\ \bibinfo {pages} {021}
  (\bibinfo {year} {2006})}\BibitemShut {NoStop}%
\bibitem [{\citenamefont {Drescher}\ \emph {et~al.}(2007)\citenamefont
  {Drescher}, \citenamefont {Dumitru}, \citenamefont {Gombeaud},\ and\
  \citenamefont {Ollitrault}}]{Drescher:2007cd}%
  \BibitemOpen
  \bibfield  {author} {\bibinfo {author} {\bibfnamefont {H.-J.}\ \bibnamefont
  {Drescher}}, \bibinfo {author} {\bibfnamefont {A.}~\bibnamefont {Dumitru}},
  \bibinfo {author} {\bibfnamefont {C.}~\bibnamefont {Gombeaud}}, \ and\
  \bibinfo {author} {\bibfnamefont {J.-Y.}\ \bibnamefont {Ollitrault}},\
  }\href@noop {} {\bibfield  {journal} {\bibinfo  {journal} {Phys. Rev.}\
  }\textbf {\bibinfo {volume} {C76}},\ \bibinfo {pages} {024905} (\bibinfo
  {year} {2007})}\BibitemShut {NoStop}%
\bibitem [{\citenamefont {Xu}\ \emph {et~al.}(2008)\citenamefont {Xu},
  \citenamefont {Greiner},\ and\ \citenamefont {Stocker}}]{Xu:2007jv}%
  \BibitemOpen
  \bibfield  {author} {\bibinfo {author} {\bibfnamefont {Z.}~\bibnamefont
  {Xu}}, \bibinfo {author} {\bibfnamefont {C.}~\bibnamefont {Greiner}}, \ and\
  \bibinfo {author} {\bibfnamefont {H.}~\bibnamefont {Stocker}},\ }\href
  {\doibase 10.1103/PhysRevLett.101.082302} {\bibfield  {journal} {\bibinfo
  {journal} {Phys. Rev. Lett.}\ }\textbf {\bibinfo {volume} {101}},\ \bibinfo
  {pages} {082302} (\bibinfo {year} {2008})}\BibitemShut {NoStop}%
\bibitem [{\citenamefont {Greco}\ \emph {et~al.}(2008)\citenamefont {Greco},
  \citenamefont {Colonna}, \citenamefont {Di~Toro},\ and\ \citenamefont
  {Ferini}}]{Greco:2008fs}%
  \BibitemOpen
  \bibfield  {author} {\bibinfo {author} {\bibfnamefont {V.}~\bibnamefont
  {Greco}}, \bibinfo {author} {\bibfnamefont {M.}~\bibnamefont {Colonna}},
  \bibinfo {author} {\bibfnamefont {M.}~\bibnamefont {Di~Toro}}, \ and\
  \bibinfo {author} {\bibfnamefont {G.}~\bibnamefont {Ferini}},\ }\href@noop {}
  {\  (\bibinfo {year} {2008})},\ \Eprint {http://arxiv.org/abs/0811.3170}
  {arXiv:0811.3170 [hep-ph]} \BibitemShut {NoStop}%
\bibitem [{\citenamefont {Lacey}\ \emph {et~al.}(2007)\citenamefont {Lacey}
  \emph {et~al.}}]{Lacey:2006bc}%
  \BibitemOpen
  \bibfield  {author} {\bibinfo {author} {\bibfnamefont {R.~A.}\ \bibnamefont
  {Lacey}} \emph {et~al.},\ }\href {\doibase 10.1103/PhysRevLett.98.092301}
  {\bibfield  {journal} {\bibinfo  {journal} {Phys. Rev. Lett.}\ }\textbf
  {\bibinfo {volume} {98}},\ \bibinfo {pages} {092301} (\bibinfo {year}
  {2007})}\BibitemShut {NoStop}%
\bibitem [{\citenamefont {Adare}\ \emph {et~al.}(2007)\citenamefont {Adare}
  \emph {et~al.}}]{Adare:2006nq}%
  \BibitemOpen
  \bibfield  {author} {\bibinfo {author} {\bibfnamefont {A.}~\bibnamefont
  {Adare}} \emph {et~al.},\ }\href@noop {} {\bibfield  {journal} {\bibinfo
  {journal} {Phys. Rev. Lett.}\ }\textbf {\bibinfo {volume} {98}},\ \bibinfo
  {pages} {172301} (\bibinfo {year} {2007})}\BibitemShut {NoStop}%
\bibitem [{\citenamefont {Luzum}\ and\ \citenamefont
  {Romatschke}(2008)}]{Luzum:2008cw}%
  \BibitemOpen
  \bibfield  {author} {\bibinfo {author} {\bibfnamefont {M.}~\bibnamefont
  {Luzum}}\ and\ \bibinfo {author} {\bibfnamefont {P.}~\bibnamefont
  {Romatschke}},\ }\href {\doibase 10.1103/PhysRevC.78.034915} {\bibfield
  {journal} {\bibinfo  {journal} {Phys. Rev.}\ }\textbf {\bibinfo {volume}
  {C78}},\ \bibinfo {pages} {034915} (\bibinfo {year} {2008})}\BibitemShut
  {NoStop}%
\bibitem [{\citenamefont {Lacey}\ \emph {et~al.}(2009)\citenamefont {Lacey},
  \citenamefont {Taranenko},\ and\ \citenamefont {Wei}}]{Lacey:2009xx}%
  \BibitemOpen
  \bibfield  {author} {\bibinfo {author} {\bibfnamefont {R.~A.}\ \bibnamefont
  {Lacey}}, \bibinfo {author} {\bibfnamefont {A.}~\bibnamefont {Taranenko}}, \
  and\ \bibinfo {author} {\bibfnamefont {R.}~\bibnamefont {Wei}},\ }\href@noop
  {} {\  (\bibinfo {year} {2009})},\ \Eprint {http://arxiv.org/abs/0905.4368}
  {arXiv:0905.4368 [nucl-ex]} \BibitemShut {NoStop}%
\bibitem [{\citenamefont {Dusling}\ \emph {et~al.}(2009)\citenamefont
  {Dusling}, \citenamefont {Moore},\ and\ \citenamefont
  {Teaney}}]{Dusling:2009df}%
  \BibitemOpen
  \bibfield  {author} {\bibinfo {author} {\bibfnamefont {K.}~\bibnamefont
  {Dusling}}, \bibinfo {author} {\bibfnamefont {G.~D.}\ \bibnamefont {Moore}},
  \ and\ \bibinfo {author} {\bibfnamefont {D.}~\bibnamefont {Teaney}},\
  }\href@noop {} {\  (\bibinfo {year} {2009})},\ \Eprint
  {http://arxiv.org/abs/0909.0754} {arXiv:0909.0754 [nucl-th]} \BibitemShut
  {NoStop}%
\bibitem [{\citenamefont {Niemi}\ \emph {et~al.}(2012)\citenamefont {Niemi},
  \citenamefont {Denicol}, \citenamefont {Holopainen},\ and\ \citenamefont
  {Huovinen}}]{Niemi:2012aj}%
  \BibitemOpen
  \bibfield  {author} {\bibinfo {author} {\bibfnamefont {H.}~\bibnamefont
  {Niemi}}, \bibinfo {author} {\bibfnamefont {G.}~\bibnamefont {Denicol}},
  \bibinfo {author} {\bibfnamefont {H.}~\bibnamefont {Holopainen}}, \ and\
  \bibinfo {author} {\bibfnamefont {P.}~\bibnamefont {Huovinen}},\ }\href@noop
  {} {\  (\bibinfo {year} {2012})},\ \Eprint {http://arxiv.org/abs/1212.1008}
  {arXiv:1212.1008 [nucl-th]} \BibitemShut {NoStop}%
\bibitem [{\citenamefont {Kovtun}\ \emph {et~al.}(2005)\citenamefont {Kovtun},
  \citenamefont {Son},\ and\ \citenamefont {Starinets}}]{Kovtun:2004de}%
  \BibitemOpen
  \bibfield  {author} {\bibinfo {author} {\bibfnamefont {P.}~\bibnamefont
  {Kovtun}}, \bibinfo {author} {\bibfnamefont {D.~T.}\ \bibnamefont {Son}}, \
  and\ \bibinfo {author} {\bibfnamefont {A.~O.}\ \bibnamefont {Starinets}},\
  }\href@noop {} {\bibfield  {journal} {\bibinfo  {journal} {Phys. Rev. Lett.}\
  }\textbf {\bibinfo {volume} {94}},\ \bibinfo {pages} {111601} (\bibinfo
  {year} {2005})},\ \Eprint {http://arxiv.org/abs/hep-th/0405231}
  {hep-th/0405231} \BibitemShut {NoStop}%
\bibitem [{\citenamefont {Schenke}\ \emph {et~al.}(2012)\citenamefont
  {Schenke}, \citenamefont {Jeon},\ and\ \citenamefont
  {Gale}}]{Schenke:2011bn}%
  \BibitemOpen
  \bibfield  {author} {\bibinfo {author} {\bibfnamefont {B.}~\bibnamefont
  {Schenke}}, \bibinfo {author} {\bibfnamefont {S.}~\bibnamefont {Jeon}}, \
  and\ \bibinfo {author} {\bibfnamefont {C.}~\bibnamefont {Gale}},\ }\href
  {\doibase 10.1103/PhysRevC.85.024901} {\bibfield  {journal} {\bibinfo
  {journal} {Phys.Rev.}\ }\textbf {\bibinfo {volume} {C85}},\ \bibinfo {pages}
  {024901} (\bibinfo {year} {2012})},\ \Eprint {http://arxiv.org/abs/1109.6289}
  {arXiv:1109.6289 [hep-ph]} \BibitemShut {NoStop}%
\bibitem [{\citenamefont {Gale}\ \emph {et~al.}(2012)\citenamefont {Gale},
  \citenamefont {Jeon}, \citenamefont {Schenke}, \citenamefont {Tribedy},\ and\
  \citenamefont {Venugopalan}}]{Gale:2012in}%
  \BibitemOpen
  \bibfield  {author} {\bibinfo {author} {\bibfnamefont {C.}~\bibnamefont
  {Gale}}, \bibinfo {author} {\bibfnamefont {S.}~\bibnamefont {Jeon}}, \bibinfo
  {author} {\bibfnamefont {B.}~\bibnamefont {Schenke}}, \bibinfo {author}
  {\bibfnamefont {P.}~\bibnamefont {Tribedy}}, \ and\ \bibinfo {author}
  {\bibfnamefont {R.}~\bibnamefont {Venugopalan}},\ }\href@noop {} {\
  (\bibinfo {year} {2012})},\ \Eprint {http://arxiv.org/abs/1210.5144}
  {arXiv:1210.5144 [hep-ph]} \BibitemShut {NoStop}%
\bibitem [{\citenamefont {Gardim}\ \emph {et~al.}(2012)\citenamefont {Gardim},
  \citenamefont {Grassi}, \citenamefont {Luzum},\ and\ \citenamefont
  {Ollitrault}}]{Gardim:2012yp}%
  \BibitemOpen
  \bibfield  {author} {\bibinfo {author} {\bibfnamefont {F.~G.}\ \bibnamefont
  {Gardim}}, \bibinfo {author} {\bibfnamefont {F.}~\bibnamefont {Grassi}},
  \bibinfo {author} {\bibfnamefont {M.}~\bibnamefont {Luzum}}, \ and\ \bibinfo
  {author} {\bibfnamefont {J.-Y.}\ \bibnamefont {Ollitrault}},\ }\href
  {\doibase 10.1103/PhysRevLett.109.202302} {\bibfield  {journal} {\bibinfo
  {journal} {Phys.Rev.Lett.}\ }\textbf {\bibinfo {volume} {109}},\ \bibinfo
  {pages} {202302} (\bibinfo {year} {2012})},\ \Eprint
  {http://arxiv.org/abs/1203.2882} {arXiv:1203.2882 [nucl-th]} \BibitemShut
  {NoStop}%
\bibitem [{\citenamefont {Luzum}\ and\ \citenamefont
  {Ollitrault}(2012)}]{Luzum:2012wu}%
  \BibitemOpen
  \bibfield  {author} {\bibinfo {author} {\bibfnamefont {M.}~\bibnamefont
  {Luzum}}\ and\ \bibinfo {author} {\bibfnamefont {J.-Y.}\ \bibnamefont
  {Ollitrault}},\ }\href@noop {} {\  (\bibinfo {year} {2012})},\ \Eprint
  {http://arxiv.org/abs/1210.6010} {arXiv:1210.6010 [nucl-th]} \BibitemShut
  {NoStop}%
\bibitem [{\citenamefont {Jia}(2011)}]{JJia:2011hfa}%
  \BibitemOpen
  \bibfield  {author} {\bibinfo {author} {\bibfnamefont {J.}~\bibnamefont
  {Jia}},\ }\href {\doibase 10.1088/0954-3899/38/12/124012} {\bibfield
  {journal} {\bibinfo  {journal} {J.Phys.}\ }\textbf {\bibinfo {volume}
  {G38}},\ \bibinfo {pages} {124012} (\bibinfo {year} {2011})},\ \Eprint
  {http://arxiv.org/abs/1107.1468} {arXiv:1107.1468 [nucl-ex]} \BibitemShut
  {NoStop}%
\bibitem [{\citenamefont {Staig}\ and\ \citenamefont
  {Shuryak}(2010)}]{Staig:2010pn}%
  \BibitemOpen
  \bibfield  {author} {\bibinfo {author} {\bibfnamefont {P.}~\bibnamefont
  {Staig}}\ and\ \bibinfo {author} {\bibfnamefont {E.}~\bibnamefont
  {Shuryak}},\ }\href@noop {} {\  (\bibinfo {year} {2010})},\ \Eprint
  {http://arxiv.org/abs/1008.3139} {arXiv:1008.3139 [nucl-th]} \BibitemShut
  {NoStop}%
\bibitem [{\citenamefont {Lacey}\ \emph {et~al.}(2011)\citenamefont {Lacey},
  \citenamefont {Taranenko}, \citenamefont {Ajitanand},\ and\ \citenamefont
  {Alexander}}]{Lacey:2011ug}%
  \BibitemOpen
  \bibfield  {author} {\bibinfo {author} {\bibfnamefont {R.~A.}\ \bibnamefont
  {Lacey}}, \bibinfo {author} {\bibfnamefont {A.}~\bibnamefont {Taranenko}},
  \bibinfo {author} {\bibfnamefont {N.}~\bibnamefont {Ajitanand}}, \ and\
  \bibinfo {author} {\bibfnamefont {J.}~\bibnamefont {Alexander}},\ }\href@noop
  {} {\  (\bibinfo {year} {2011})},\ \Eprint {http://arxiv.org/abs/1105.3782}
  {arXiv:1105.3782 [nucl-ex]} \BibitemShut {NoStop}%
\bibitem [{\citenamefont {Lacey}\ \emph {et~al.}(2010)\citenamefont {Lacey},
  \citenamefont {Wei}, \citenamefont {Ajitanand},\ and\ \citenamefont
  {Taranenko}}]{Lacey:2010hw}%
  \BibitemOpen
  \bibfield  {author} {\bibinfo {author} {\bibfnamefont {R.~A.}\ \bibnamefont
  {Lacey}}, \bibinfo {author} {\bibfnamefont {R.}~\bibnamefont {Wei}}, \bibinfo
  {author} {\bibfnamefont {N.~N.}\ \bibnamefont {Ajitanand}}, \ and\ \bibinfo
  {author} {\bibfnamefont {A.}~\bibnamefont {Taranenko}},\ }\href@noop {} {\
  (\bibinfo {year} {2010})},\ \Eprint {http://arxiv.org/abs/1009.5230}
  {arXiv:1009.5230 [nucl-ex]} \BibitemShut {NoStop}%
\bibitem [{\citenamefont {Drescher}\ and\ \citenamefont
  {Nara}(2007)}]{Drescher:2007ax}%
  \BibitemOpen
  \bibfield  {author} {\bibinfo {author} {\bibfnamefont {H.-J.}\ \bibnamefont
  {Drescher}}\ and\ \bibinfo {author} {\bibfnamefont {Y.}~\bibnamefont
  {Nara}},\ }\href {\doibase 10.1103/PhysRevC.76.041903} {\bibfield  {journal}
  {\bibinfo  {journal} {Phys. Rev.}\ }\textbf {\bibinfo {volume} {C76}},\
  \bibinfo {pages} {041903} (\bibinfo {year} {2007})}\BibitemShut {NoStop}%
\bibitem [{\citenamefont {Kharzeev}\ and\ \citenamefont
  {Nardi}(2001)}]{Kharzeev:2000ph}%
  \BibitemOpen
  \bibfield  {author} {\bibinfo {author} {\bibfnamefont {D.}~\bibnamefont
  {Kharzeev}}\ and\ \bibinfo {author} {\bibfnamefont {M.}~\bibnamefont
  {Nardi}},\ }\href {\doibase 10.1016/S0370-2693(01)00457-9} {\bibfield
  {journal} {\bibinfo  {journal} {Phys.Lett.}\ }\textbf {\bibinfo {volume}
  {B507}},\ \bibinfo {pages} {121} (\bibinfo {year} {2001})},\ \Eprint
  {http://arxiv.org/abs/nucl-th/0012025} {arXiv:nucl-th/0012025 [nucl-th]}
  \BibitemShut {NoStop}%
\bibitem [{\citenamefont {Lappi}\ and\ \citenamefont
  {Venugopalan}(2006)}]{Lappi:2006xc}%
  \BibitemOpen
  \bibfield  {author} {\bibinfo {author} {\bibfnamefont {T.}~\bibnamefont
  {Lappi}}\ and\ \bibinfo {author} {\bibfnamefont {R.}~\bibnamefont
  {Venugopalan}},\ }\href {\doibase 10.1103/PhysRevC.74.054905} {\bibfield
  {journal} {\bibinfo  {journal} {Phys. Rev.}\ }\textbf {\bibinfo {volume}
  {C74}},\ \bibinfo {pages} {054905} (\bibinfo {year} {2006})}\BibitemShut
  {NoStop}%
\bibitem [{\citenamefont {Miller}\ \emph {et~al.}(2007)\citenamefont {Miller},
  \citenamefont {Reygers}, \citenamefont {Sanders},\ and\ \citenamefont
  {Steinberg}}]{Miller:2007ri}%
  \BibitemOpen
  \bibfield  {author} {\bibinfo {author} {\bibfnamefont {M.~L.}\ \bibnamefont
  {Miller}}, \bibinfo {author} {\bibfnamefont {K.}~\bibnamefont {Reygers}},
  \bibinfo {author} {\bibfnamefont {S.~J.}\ \bibnamefont {Sanders}}, \ and\
  \bibinfo {author} {\bibfnamefont {P.}~\bibnamefont {Steinberg}},\ }\href
  {\doibase 10.1146/annurev.nucl.57.090506.123020} {\bibfield  {journal}
  {\bibinfo  {journal} {Ann. Rev. Nucl. Part. Sci.}\ }\textbf {\bibinfo
  {volume} {57}},\ \bibinfo {pages} {205} (\bibinfo {year} {2007})}\BibitemShut
  {NoStop}%
\bibitem [{cms()}]{cms_ulc_note}%
  \BibitemOpen
  \href@noop {} {}\bibinfo {note} {See Fig. 14 in CMS PAS
  HIN-12-011.}\BibitemShut {Stop}%
\bibitem [{\citenamefont {Schenke}\ \emph {et~al.}(2011)\citenamefont
  {Schenke}, \citenamefont {Jeon},\ and\ \citenamefont
  {Gale}}]{Schenke:2011tv}%
  \BibitemOpen
  \bibfield  {author} {\bibinfo {author} {\bibfnamefont {B.}~\bibnamefont
  {Schenke}}, \bibinfo {author} {\bibfnamefont {S.}~\bibnamefont {Jeon}}, \
  and\ \bibinfo {author} {\bibfnamefont {C.}~\bibnamefont {Gale}},\ }\href
  {\doibase 10.1016/j.physletb.2011.06.065} {\bibfield  {journal} {\bibinfo
  {journal} {Phys.Lett.}\ }\textbf {\bibinfo {volume} {B702}},\ \bibinfo
  {pages} {59} (\bibinfo {year} {2011})},\ \Eprint
  {http://arxiv.org/abs/1102.0575} {arXiv:1102.0575 [hep-ph]} \BibitemShut
  {NoStop}%
\bibitem [{\citenamefont {Qiu}\ \emph {et~al.}(2012)\citenamefont {Qiu},
  \citenamefont {Shen},\ and\ \citenamefont {Heinz}}]{Qiu:2011hf}%
  \BibitemOpen
  \bibfield  {author} {\bibinfo {author} {\bibfnamefont {Z.}~\bibnamefont
  {Qiu}}, \bibinfo {author} {\bibfnamefont {C.}~\bibnamefont {Shen}}, \ and\
  \bibinfo {author} {\bibfnamefont {U.}~\bibnamefont {Heinz}},\ }\href
  {\doibase 10.1016/j.physletb.2011.12.041} {\bibfield  {journal} {\bibinfo
  {journal} {Phys.Lett.}\ }\textbf {\bibinfo {volume} {B707}},\ \bibinfo
  {pages} {151} (\bibinfo {year} {2012})},\ \Eprint
  {http://arxiv.org/abs/1110.3033} {arXiv:1110.3033 [nucl-th]} \BibitemShut
  {NoStop}%
\end{thebibliography}%
\end{document}